\documentstyle[aps]{revtex}
\begin{document}
\title
{\bf
Nagaoka  State and Degeneracy in the $U=\infty$ Hubbard Model
}
\author{
V. Subrahmanyam and Fajian Shi}
\address{
Max-Planck Institut F\"ur Physik Komplexer Systeme,
Bayreuther Str. 40, D-01187 Dresden, Germany.\\
\parbox[t]{14 truecm}{\small
The $U=\infty$ Hubbard model on a hypercubic lattice with exactly one hole is
studied. It is seen that a degeneracy can occur in the maximal-spin
sector. It is proved variationally that 
the Nagaoka (maximal spin) state is not a ground state, if degeneracy
exists.  We give an explicit construction of a variational
state, an incommensurate spin-wave state, with energy lower than that of the 
Nagaoka state. For the case of more than one hole, a few examples are discussed
where a degeneracy exists.
}}
\maketitle
\vskip 1cm

The Hubbard model plays a key role in the study of magnetism, metal-insulator
transitions, and high-temperature superconductivity. It is a simple model that
captures the physics of phenomena of correlated systems. In view of the 
difficulty in establishing rigorous results of the model, the Nagaoka's theorem
\cite{1}
is very significant, providing a good example, though of a restricted use, of
metallic ferromagnetism in itinerant electron systems. The result of Nagaoka
that
the ground state of the Hubbard model with an infinite on-site repulsion with
exactly one hole is a maximal-spin state is remarkable, given that for 
large on-site repulsion with no holes a maximal-spin state cannot be a 
ground state, due to the antiferromagnetic correlations that arise from kinetic
exchange phenomenon. There have been several contributions extending or 
simplifying the original result of Nagaoka
\cite{2,3,4}, and investigations on the instability
of the Nagaoka state, when the number of holes is more than
one\cite{5,6,7}. 
In the following we give a simple proof that the  Nagaoka state is not a
ground state if there is a degeneracy in the
maximal-spin sector\cite{7}.
We follow a variational construction of a spin-wave state,
and in conjunction with the fact that a degeneracy exists, and show 
that by lowering the spin
one can lower the energy below that of the Nagaoka state. The simplest case 
where
a degeneracy occurs is the one-hole case for a closed chain with an odd
number of sites, and
similarly the higher-dimensional lattices with an odd number of sites in at
least one direction.

Let us consider the Hubbard model on a hypercubic lattice with $N$ sites ($N_1 
\times N_2 \times N_3..$, with $N_i$ number of sites in the $i$'th direction), 
and $z$ coordination number, with periodic boundary conditions. The
Hamiltonian can be written as
\begin{equation}
H=-t\eta\sum_{<ij>\sigma}c^\dagger_{i\sigma}c_{j\sigma}+ U\sum_{i}
n_{i\uparrow}n_{i\downarrow}
\end{equation} 
where $c_{i\sigma}(c^\dagger_{i\sigma})$ annihilates (creates) an electron of
spin $\sigma$ at $i$'th lattice site, and $n_{i\sigma}=c_{i\sigma}^\dagger c_{
i\sigma}$ is the number operator. The first sum in the above equation is
for nearest neighbours only, with the hopping strength $t$. The parameter $
\eta=1$ for the case of fermions (that we will discuss through out), and $\eta
=-1$ for hard-core bosons. All the conclusions that we draw for the fermion
problem carry through to hard-core bosons, by changing the sign of $\eta$.
We consider the
case of the on-site repulsion $U$ to be infinitely large. This forces all the
physical states with any number of electrons to consist of no doubly-occupied
site orbitals, tantamounting to a super-Pauli exclusion. We choose a basis
set of states, a $3^N$-dimensional Hilbert space, with this constraint built in
and consequently the
Hamiltonian consists of only the first term in the above equation. 
The total spin $S$, and the z-component of the
total spin $S^z$ (at $i$'th site the spin operators are given as $S_i^z=(n_{
i\uparrow}-n_{i\downarrow})/2, 
S_i^-=c_{i\downarrow}^{\dagger} c_{i\uparrow}$) are good quantum numbers of
the model, and we use them to label the sectors of the Hilbert space. 
Let the number of electrons $N_e=N-1$. 
The basis states consist of direct 
products of
$N-1$ site orbitals occupied each with either a spin-up or down electron. 
Also we note that all the matrix elements of the Hamiltonian
have a definite sign, by choosing the set of basis states carefully.
In the sector with $S^z=S_{max}=N_e/2$, the
basis states can be labeled by the location of hole, and are obtained by
operating with an annihilation operator on a reference ferromagnetic state 
(no holes and no down spins). 
The basis state $i$ is given by
\begin{equation}
|i>=c_{i\uparrow}|{\uparrow\uparrow\uparrow\uparrow..}>\equiv c_{i\uparrow}|REF>
\end{equation}
where the hole is at the $i$'th site.
The reference state $|REF>=\prod_{i=1,N} c_{i\uparrow}^\dagger |{0}>$ is 
created from
the vacuum state by using the creation operators.
With this choice of the basis set,
the matrix elements have all one same sign, $<i|H|j>=t\zeta_{i,j}$ where
$\zeta_{i,j}=1$ if the site $i$ and $j$ are nearest neighbours on the
lattice, and zero otherwise. For the sectors with $S^z\ne S_{max}$ the 
construction is similar.
For instance, for $S^z=S_{max}-1$ sector, the basis states are $\{|ik>\}$. A
basis state is
labeled by the locations of the hole (at $i$) and the down spin (at $k$), and
is constructed from the reference state as $|ik>=c_{i\uparrow}c_{k\downarrow}^{
\dagger}c_{k\uparrow}|REF>$.
The overlaps between the basis states are all of
the same sign (using the fermion anticommutation relations),
\begin{equation}
<i^\prime k^\prime| H |ik> = t \zeta_{i,i^\prime} ( \delta_{k,k^\prime} + \delta_{
i,k^\prime}\delta_{i^\prime,k})
\end{equation}
where the first term in the parenthesis comes from a movement of the up-spin 
electron, and the
second term from that of the down-spin electron. It should be noted that
the above equation does not change for hard-core bosons (by using the boson
commutation relations), if we use $\eta=-1$ in Eq.(1).

As can be seen from the above equation, the Hamiltonian in a matrix form looks
like $H=t A$, in any $S^z$ sector, where $A$ is matrix with
positive-semidefinite elements, $A_{ij}\ge 0$, and the size of the matrix 
depends on $S^z$. The diagonal elements are zero, and there are exactly $z$
nonzero elements on any row (of magnitude unity). The largest eigenvalue is 
$\lambda_{max}\le z$ independent of size, from Gershgorin theorem. And its easy
to check that a uniform vector (which is a maximal-spin state) is eigenstate of
the matrix, with eigenvalue $z$, implying $\lambda_{max}=z$. If the matrix is
'connected' \cite{2,3}in the configuration space ($i.e.$ the 
elements of $B=A^M$, 
for a finite $M$, are positive definite, $B_{ij}>0$), $\lambda_{max}$ is a 
non-degenerate eigenvalue for any finite size from Perron-Frobenius theorem. 
For a bipartite lattice, where
$N_1, N_2..$ are all even, the sign of $t$ is irrelevant. This can be seen by
a canonical transform, changing the sign of the electron creation and 
annihilation operators on one of the two sublattices. It suffices to study one
sign of $t$. So for $t<0$, the lowest eigenvalue of the Hamiltonian corresponds
to the highest eigenvalue of the matrix $A$. In this case,
the Nagaoka state ($S=S_{max}$) is a ground state, with an energy $- z|t|$,
and also it is nondegenerate (connectivity of $A$ is provable in more 
than one dimension)\cite{3}). For
nonbipartite lattices, where at least one of $N_1,N_2..$ is an odd number, we
have to study both signs of $t$, and the nature of the ground state is quite
different, as we shall see below. We proceed for $t<0$ similarly as above, as 
the
problem boils down to the investigation of $\lambda_{max}$, and
the Nagaoka state is a ground state. The more interesting case, 
however, is for $t>0$, where a degeneracy exists in the maximal-spin sector.
Here we are confronted with the lowest eigenvalue 
$\lambda_{min}$ of $A$
and it very well could depend on the size of $A$; all the sectors of $S^z$ have 
to be examined. 
Table 1 shows a few examples of periodic two-dimensional lattices where a 
degeneracy occurs, along with the
lowest eigenvalues from numerical diagonalizations for two spin sectors
$S=S_{max}$ and $S_{max}-1$, and the Nagoaka state fails to be a ground
state.
It can be shown from a variational construction that degeneracy 
disfavors high-spin states\cite{7}.
We give a simple proof based on a spin-wave state that the  
Nagaoka state is not a ground state, as a direct consequence of a
degeneracy. Our variational state has simple structure, and
is similar to Roth-type wavefunctions investigated earlier
\cite{6,7}.
The strategy is to look at two sectors
with $S^z=S_{max}$ and $S^z=S_{max}-1$, and show that $\lambda_{min}(S_{max}-1)
< \lambda_{min}(
S_{max})$, which implies that a state with total spin $S=S_{max}-1$ has lower
energy than the energy of Nagaoka state. 

We focus on a one-dimensional closed chain with an odd-number of sites, and
all the arguments used below can be trivially extended to higher-dimensional
systems. In the sector with $S^z=S_{max}$ we have only one species of electrons,
which makes the system a spinless fermion problem. The term with the coulomb 
interaction $U=\infty$ in Eq. (1) does not come into play at all, and all the
many-particle eigenfunctions are constructed as direct products of the 
single-fermion states. The translational invariance of the lattice helps us
in diagonalizing the Hamiltonian in momentum $k$ space. The energy spectrum (the
many-electron spectrum or the one-hole spectrum)
is given by $\varepsilon_k=2t\cos{k}$, where $k=2\pi n/N$ and $n$ an integer.
For $t>0$ (and setting $t=1$ from now onwards), the lowest
energy of the spectrum is for $k=-\pi\pm \pi/N$ with a two-fold degeneracy (and
the two degenerate states have a net momentum), and the highest energy is for
$k=0$ which is non-degenerate. Similarly for a square lattice 
with $N_1
\times
N_2$ sites and $N_1$ odd and $N_2$ even, the lowest-energy state is two-fold
degenerate. In this sector all the states carry a total spin of $S=S_{max}$, 
and hence this is an orbital degeneracy. Let us denote the two degenerate 
lowest-energy states in this sector as
\begin{equation}
\phi_1=\sum a_i |i>; \phi_2 =\sum b_i |i>.
\end{equation}
We have $H|\phi_{1,2}>=E_M|\phi_{1,2}>$, where the energy is $E_M=-2\cos{\pi/N}
$, and the following conditions on the eigenfunctions from normalization and
orthogonality 
\begin{equation}
\sum a_i^\star a_i=
\sum b_i^\star b_i =1; \sum a_i^\star b_i=0;
\end{equation}
and further since the two states are eigenstates of the Hamiltonian, we have
\begin{equation}
\sum_{<ij>} a_i^\star a_j=\sum_{<ij>} b_i^\star b_j=E_M; \sum_{<ij>}a_i^\star 
b_j =0.
\end{equation}
More importantly the two degenerate eigenfunctions are related by a phase
\begin{equation}
a_j=b_j e^{i q_o r_j}
\end{equation}
where $q_o=2\pi/N$. As we shall see later this fact will be very crucial and
useful.

We now turn to the sector with $S^z=S_{max}-1$, where an eigenstate of the
Hamiltonian will either carry a total spin $S=S_{max}$ or $S=S_{max}-1$, as
no other spin is possible. To diagonalize the Hamiltonian in this sector
is a nontrivial problem as we have one down spin this time. However, we may
not be impeded by that. The normalized eigenstates with $S=S_{max}$ are easy to
construct as the Hamiltonian commutes with the total spin, so a lowering of the
z-component of the states we constructed above gives us the eigenstates we
desire here. The two maximal-spin lowest-energy degenerate states are
$S^-\phi_{1,2}$, with an energy $E_M$. The eigenstates with $S=S_{
max}-1$ are not easy to construct. We resort to a variational method.
Let us construct a general
variational spin-wave state
\begin{equation}
\phi_a(q)={1\over\sqrt{N-1}} \sum_i \sum_{k\ne i} a_i e^{i q r_k} |ik>
\end{equation}
where the basis state $|ik>$ has a hole at the site $i$ and a down spin at $k$,
as discussed earlier. It is clear, from Eq.(5), that the above state
is normalized. For $q=0$ the state is an eigenstate of the 
Hamiltonian with maximal spin. 
For $q\ne 0$, the state does not have a definite spin,
the expectation value of the spin is less than $S_{max}$, as it has a component
from both $S_{max}$ and $S_{max}-1$ spin sectors. If $q\ne 2\pi n/N$, with an
integer $n$, the above is an incommensurate spin-wave state. It is 
straightforward to evaluate the variational energy of this state using Eqs.
(3) and (6), and we get
\begin{equation}
E_a(q)=<\phi_a(q)|H |\phi_a(q)>=E_M +{A(q)-E_M\over N-1}
\end{equation}
where the fourier-transform function
\begin{equation}
A(q)=\sum_{<ij>} a_i^\star a_j e^{iq(r_i-r_j)}.
\end{equation}
It is easy to show that $A(q)\ge E_M$, by using a
variational principle on a state $\phi=\sum a_i \exp({i q r_i})|i>$ in 
the $S^z=S_{max}$ sector, where $\phi_1$ (see Eq.(4)) is a ground state. This 
implies $E_M \le E_a(q)$. The equality holds at
two points $q=0$ and $q=q_o$, as we will see below. From a similar construction
of a spin-wave state $\phi_b(q)$, using $b_i$ instead of $a_i$ in Eq.(8),
we get the variational energy $E_b(q)=E_M +[B(q)-E_M]/N-1$, where
\begin{equation}
B(q)=\sum_{<ij>}b_i^\star b_j e^{i q(r_i-r_j)}.
\end{equation}
The two fourier-transform functions are related as the two degenerate states
are related by a phase (see Eq.(7)); we have
\begin{equation}
A(q)=B(q-q_o).
\end{equation} 
The above equation, along with Eqs.(6), (10) and (11), implies
$A(q_o)=B(0)=E_M=A(0)=B(-q_o)$. We have
now two different spin-wave states constructed from the two degenerate 
lowest-energy states with $S=S_{max}$. Though both these states have energies
higher than $E_M$, we can now superpose these states to hope for a lowering
in energy, as their overlap could be nonzero and useful. A precaution, that
we have 
to take, is that the variational state should not become a spin eigenstate with 
$S=S_{max}$.
Let us try a variational state 
\begin{equation}
\psi={1\over 2} (\phi_a(q)-\phi_b(q^\prime))
\end{equation}
with a choice, from hindsight, $q^\prime=q-q_o$. Though we are restricting the
variational parameter space, we shall see below that we still have enough 
freedom to minimize the variational energy below $E_M$. 

The above variational state has a norm $<\psi|\psi>=N/N-1$, and the overlap
between the two components of the state for the choice $q^\prime=q-q_o$ is 
$<\phi_b(q^{
\prime})|H|\phi_a(q)>= B(q)-B(q_o)-E_M/N-1$, where we have used Eqs. (3),(7)
and (11). The variational energy $E_\psi=<\psi|H|\psi>/<\psi|\psi>$ is, using 
the expressions for $E_a(q)$, and $E_b(q^\prime)$ from Eq. (9),
\begin{equation}
E_\psi = E_M +{1\over 2N}( B(q+q_o)+B(q-q_0)-2B(q))
\end{equation}
where we used Eq.(12) to write all the fourier-transform functions in terms of
$B(q)$. Now we seek values of $q$ such that $\gamma (q)\equiv
B(q+q_o)+B(q-q_0)-2B(q)
$ is negative, which
implies $E_\psi < E_M$. We note that the function $B(q)$ is periodic (with
period $2\pi$), and is non-constant, $B(0)=B(-q_o)=E_M<0$, and $B(\pi)=-E_M$. 
Then it is guaranteed that $\gamma (q) < 0$, for some value of $q$,
as the function $B(q)$ necessarily has a local
maximum. The expression for $\gamma (q)$ can be substantially simplified using
Eq. (11) as $\gamma (q) = -2 (1-\cos {q_o}) B(q)$. Now any value of $q$
for $B(q) >0$ would suffice to give us the desired result $\gamma (q) <0$. For
$q=\pi$, we have $\gamma (\pi)= -4(1-\cos {q_o})\cos{(\pi/N)}$, and
\begin{equation}
E_\psi = E_M +{2\over N}(1-\cos {q_o})\cos{\pi\over N} < E_M.
\end{equation}
This completes the proof that the variational state in Eq.(13), with $q=\pi,
q^\prime=\pi-q_o$, has lower energy than that of the Nagaoka ($S=S_{max}$)
state. The fact that we have two degenerate lowest-energy states with maximal
spin is very crucial. The above arguments can be readily extended for more than
one dimension. The structure of the proof does not change for 
higher-dimensional systems. For instance consider a two-dimensional lattice with
periodic boundary conditions, with N=$N_1\times N_2$ sites. And let $N_1$ be
an even number and $N_2$  odd. The energy spectrum
for the hole in $S^z=S_{max}$ sector is $\varepsilon(k_1,k_2)=2(\cos k_1 + \cos
{k_2})$, with $k_i=2\pi n_i/N_i$. The lowest energy is for $k_1=\pi, k_2 =
\pi \pm \pi/N_2$, with $E_M= -2(1+\cos({\pi/N_2}))$, with a two-fold 
degenerate.
The various quantities $q,q_o,\{r_i\}$ now become two-dimensional vectors. The 
phase
between the two degenerate eigenstates is, in Eq.(7), is ${\vec q}_0=(0,2\pi/
N_2)$ and we choose ${\vec q}=(0,\pi)$ in Eq. (14), to get the desired result
$E_\psi < E_M$. In a general case of a hypercubic lattice with $N=N_1\times
N_2\times N_3...$ sites, and with at least one of $N_i$ is an odd number, we
get a two-fold degenerate maximal-spin state for $t>0$, where the Nagaoka state
is not a ground state. We expect that a similar proof can be constructed for
any lattice with exactly one hole, when a degeneracy exists.

We would like to briefly comment on the
nature of the actual ground state, if degeneracy exists. We expect
similar arguments can be constructed to show that the lower the spin the better
is the ground state energy, and most probably the ground state has the
lowest possible spin. 
A heuristic argument can be given to understand 
the trend of minimal-spin ground state when degeneracy exists.
In the non-degenerate case, the ground
state eigenfunction is positive definite ($i.e.$ a basis can be chosen such
that the probability amplitudes of all the basis states in the ground state
are positive), which is the key point of the Nagaoka state. If we have 
a degeneracy, the orthogonal states cannot 
be both positive definite, and at least one of them have negative probability
amplitudes. A lowering of the spin introduces, by itself, negative signs. For
instance, any state with $S=S^z=S_{max}-1$ is of the form
$|S=S^z=S_{max}-1>=\sum_{ik} \chi_{ik} |ik>$.
It is necessarily accompanied by the constraints (as $S^+|S=S^z=S_{max}-1>=0$)
$\sum_{k} \chi_{ik} =0 ~~~{\rm for~~a~~fixed}~~i.$
This implies that the amplitudes come with both signs to satisfy the constraint
equations. Now it may be expected that these negative signs, and the negative
signs as a consequence of degeneracy may conspire to cancel out making
the eigenfunction as much 'positive definite' as possible, and translate into
a kinetic energy lowering. And by lowering the
spin further, one lowers the energy further. Hence an expected minimal spin
ground state.
 
We discuss the case when the number of holes $N_H$ is more than one. It
is easy to check if the maximal-spin sector has a degeneracy in the
lowest many-hole energy state, as all we need to solve is a spinless fermion 
problem. We have the following interesting and simple examples where
degeneracy occurs.
(a) Consider the case of a bipartite lattice where for $N_H=1$
we have no degeneracy and a Nagaoka ground state. For
a closed chain with an even number of sites, the one-hole energy
spectrum in $S^z=S_{max}$ sector is $\varepsilon (k)= -2|t| \cos{k}$. When
we have two holes, the ground state in this sector is a direct product of
single-hole $k-$space eigenstates (not true for hard-core bosons) with $k_1=0,
k_2=\pm 2\pi/N$, which is
two-fold degenerate. The same holds for $N_H$ even and less than
$N-1$. And similarly for
an $N_1\times N_2$ lattice, for many values of $N_H$ the lowest-energy state
in the maximal-spin sector is degenerate. In the particular case of quarter
filling $N_e=N_H=N/4$, for any $N$, there is a degeneracy.
(b) Consider the case
of a non-bipartite lattice, with degeneracy for $N_H=1$ (which implies
Nagaoka is not a ground state from the proof outlined above). 
For example a closed chain with an odd number of sites has a degeneracy
in the lowest-energy state of the maximal-spin sector when 
$N_H$ is odd and less than $N-1$. Similarly for higher-dimensional systems we
encounter degeneracy for many values of $N_H$. 
(c) Consider the case of a long-range hopping model, where the constraint of 
nearest-neighbour-only hopping that we have in Eq.(1) is relaxed, which
exhibits macroscopic degeneracy. It is easy to 
diagonalize the maximal-spin sector, and the one-hole spectrum has only
two energy levels, with energies $-t(N-1)$ (non-degenerate), and $-t$ 
(degeneracy $N-1)$.  For $t>0$, the lowest-energy many-hole
state for $N_H<N-1$ is degenerate. For $t<0$, there is degeneracy
for $1<N_H <N-1$. The singular case of $N_H=N-1$ also has a degenerate
lowest-energy state, but there is only one spin sector $S=1/2$. 

In all the examples we considered above there is a degeneracy 
in the lowest-energy state of maximal-spin sector. We expect that the Nagaoka
state is not a ground state in these cases. It would be 
interesting to see if one can generalize our simple proof for the above 
examples\cite{7}. 
Particularly, it should be investigated whether the ingredients that we used for
$N_H=1$, $viz.$ the translational
invariance, which is important to diagonalize the Hamiltonian in $S^z=S_{max}$
sector and get a handle on the degenerate states (Eq.(7)), and the
choice of a basis states such that all the overlaps have a single sign (Eq.(3)),
are essential.  Currently an investigation
is under progress to extend our procedure in a more general context, $viz.$
for non-cubic lattices with more than one hole.

In conclusion, we have seen that there is a general trend in the nature of
the ground state. For $N_H=1$, when the maximal-spin sector has a non-degenerate
lowest-energy state, the ground state is a maximal-spin Nagaoka state, and
a degeneracy disfavors higher-spin states and a minimal-spin ground
state is expected. An explicit proof,  
based on a variational construction of
a spin-wave state, is given to exhibit that the Nagaoka state is not a ground
state
if degeneracy exists on hybercubic lattices, with an odd number of
sites in at least one direction. 

We would like to thank K. Hallberg for many useful discussions, and K. Penc
for bringing the work of Suto\cite{7} to our attention.

\vskip 0.4 cm
\noindent {\bf Table 1} The lowest eigenvalue $E(S)$ for two spin sectors $S=S_{max},
S_{max}-1$ for various sizes for $t=1$.
\vskip 0.4cm
\begin{tabular} {|c|c|c|c|}
\hline
$N_1, N_2$& Degeneracy in &$E(S_{max})$& $E(S_{max}-1)$\\
& $S_{max}$ Sector& &\\
\hline
3,3&4&-2.0&-2.732\\
\hline
3,4&2&-3.0&-3.342\\
\hline
3,5&4&-2.618&-3.115\\
\hline
4,4&1&-4.0&-3.924\\
\hline
3,6&2&-3.0&-3.278\\
\hline
4,5&2&-3.618&-3.713\\
\hline
\end{tabular}

\begin{thebibliography}{99}
\bibitem{1}Y. Nagaoka, Phys. Rev. {\bf 147}, 392(1966).
\bibitem{2}H. Tasaki, Phys. Rev. {\bf B40}, 9192 (1989).
\bibitem{3}G. Tian, J. Phys. {\bf A23}, 2231 (1990).
\bibitem{4}E. Lieb, Phys. Rev. Lett. (1989)
\bibitem{5} B. Doucot and X. G. Wen, Phys. Rev. {\bf 40}, 2719 (1989).
\bibitem{6}B. S. Shastry, H. R. Krishnamurthy and P. W. Anderson, Phys.
Rev. {\bf B41}, 2375 (1990).
\bibitem{7} A. Suto, Commun. Math. Phys. {\bf 140}, 43 (1991).
\end{thebibliography}
\end{document}